\documentclass[aps,pra,twocolumn,10pt,showpacs,amsmath,amssymb,floatfix,superscriptaddress,longbibliography]{revtex4-1}
\usepackage{graphicx}
\usepackage{amsmath}
\usepackage{amssymb}
\usepackage{nccmath}
\usepackage[utf8]{inputenc}
\usepackage{comment}
\usepackage{dcolumn}
\usepackage{bm}
\usepackage{hyperref}
\usepackage{color}
\usepackage{soul}

\newcommand{\mP}{\mathcal{P}}

\newcommand{\Rmnum}[1]{\expandafter\@slowromancap\romannumeral  #1@}

\newcommand{\Hxc}{\textrm{Hxc}}
\newcommand{\br}{\mathbf{r}}

\newcommand{\hH}{\hat{H}}
\newcommand{\hT}{\hat{T}}
\newcommand{\hV}{\hat{V}}
\newcommand{\hW}{\hat{W}}

\newcommand{\half}{\frac{1}{2}}

\newcommand{\Eq}[1]{Eq.(\ref{#1})}

\newcommand{\Fig}[1]{FIG. \ref{#1}}

\usepackage{bbold}

\newcommand{\ssc}[1]{{\scriptscriptstyle #1}}

\begin{document}

\title{Strictly correlated electrons approach to excitation energies of dissociating molecules}

\author{Luis Cort} 
\affiliation{Department of Physics,
Nanoscience Center P.O.Box 35 FI-40014 University of Jyv\"{a}skyl\"{a}, Finland}

\author{Soeren Ersbak Bang Nielsen} 
\affiliation{Max Planck Institute for the Structure and Dynamics of Matter and Center for Free-Electron Laser Science, Luruper Chaussee 149, 22761 Hamburg, Germany}

\author{Robert van Leeuwen}
\affiliation{Department of Physics,
Nanoscience Center P.O.Box 35 FI-40014 University of Jyv\"{a}skyl\"{a}, Finland}

\begin{abstract}
In this work we consider a numerically solvable model of a two-electron diatomic molecule to study a recently proposed approximation based on the density-functional theory of so-called strictly correlated electrons (SCE).
We map out the full two-particle wave function for a wide range of bond distances and interaction strengths
and obtain analytic results for the two-particle states and eigenenergies in various limits of strong and weak interactions, and in the limit of large bond distance. 
We then study the so-called Hartree-exchange-correlation (Hxc) kernel of time-dependent density functional theory which is a key ingredient in calculating excitation energies.
We study an approximation based on adiabatic SCE (ASCE) theory 
which was shown to display a particular feature of the exact Hxc-kernel, namely a spatial divergence as function of the bond distance. This makes the ASCE kernel a candidate for correcting
a notorious failure of the commonly used adiabatic local density approximation (ALDA) in the calculation of excitation energies of dissociating molecules. Unlike the ALDA, we obtain non-zero excitation energies 
from the ASCE kernel in the dissociation regime but they do not correspond to those of the true spectrum unless the interaction strength is taken to be very large such that the SCE theory has the right regime of validity,
in which case the excitation energies become exact and represent the so-called zero point oscillations of the strictly correlated electrons.
The commonly studied physical dissociation regime, namely large molecular separation at intermediate interaction strength, therefore remains a challenge for density functional approximations based on SCE theory.   
\end{abstract}

\maketitle
\section{\textbf{Introduction}}
Density-functional theory (DFT) is a commonly used electronic structure method. Its ground state version is mainly used to calculate energies and structures of electronic systems \cite{Barth:review}, while its time-dependent (TD) counterpart TDDFT also
allows for the calculation of dynamic properties and excitation energies  \cite{Ullrich:book}.
Virtually all density-functional calculations are based on the Kohn-Sham (KS) system, a non-interacting system that produces the same electronic density as the true system of interest.
The KS system provides a considerable simplification of the many-body problem which is advantageous for numerical implementations. However, all the complications of the true many-body system are
hidden in the effective potential of the KS-system.
This KS potential is typically expressed as a sum of the external potential of the interacting system of interest and the Hartree-exchange-correlation (Hxc) potential containing implicitly the many-body effects of the interacting system.
The KS formalism is equally applicable in ground state and time-dependent DFT but in this work we will focus on the calculation of excitation energies
which are obtained in TDDFT using a linear response formalism. 
For this purpose, it is enough to know the functional derivative of the Hxc potential with respect to the density which yields a quantity known as the Hxc-kernel. 
The simplest possible approximation for the Hxc kernel is the adiabatic local-density approximation (ALDA), for which the kernel is local in space and time. 
Although this approximation has been used successfully~\cite{Ullrich:book} it has a number of important deficiencies, such as
the inability to reproduce Born-Oppenheimer surfaces of excited states in dissociating molecules~\cite{GritsenkoBaerends, GritsenkoBaerends2}. 

When a molecule separates into fragments
its excitation energies should approach those of the separate fragments. This behavior is not reproduced by the ALDA since
upon dissociation the gap between the bonding and anti-bonding KS eigenvalues decreases exponentially fast with the bond distance,
and the ALDA kernel is unable to correct for this thereby rendering many of the excitation energies to become zero in the dissociation limit.
To correct for this, asymptotic corrections have been devised \cite{GritsenkoBaerends, GritsenkoBaerends2}
 that introduce exponentially growing terms in the kernel that compensate for the closing of the bonding-antibonding gap.
Although such corrections can reproduce the main features of the exact bonding curve for the lowest excited state \cite{GritsenkoBaerends, GritsenkoBaerends2}, there is no systematic way to construct such functionals. Other more systematic approximations often rely on perturbative expansions, which  makes them questionable in the multi-configuration regime required to describe molecular dissociation.

In recent work~\cite{Lani} an approximate kernel was derived within the framework of so-called strictly correlate electrons (SCE). This is a ground state DFT formalism that becomes exact in the limit of very large two-body interactions.
When the simplest approximation within this formalism is applied within the adiabatic approximation an approximate Hxc kernel can be derived . This so-called adiabatic SCE (ASCE) kernel was shown to have a number of desirable
features. It was shown to obey the so-called zero-force theorem~\cite{Ullrich:book,Mundt:PRA} and it was shown that in the case of molecular dissociation it exhibits an exponential growth with the bond distance~\cite{Lani}. The kernel therefore displays
a very non-local spatial behavior that has the potential to cure the deficiency of the ALDA kernel for molecular dissociation.
We recently investigated the ASCE  kernel \cite{Cort} in a model system for which the exchange-correlation kernel can be obtained exactly for various two-body interaction strengths.
It was found that the leading order and the next to leading order of the asymptotic expansion for the exact Hxc kernel in terms of the interaction strength agreed with that one predicted by the adiabatic SCE formalism. This result shows that the SCE formalism is a promising method for describing the linear response properties in the strong interaction limit. 
Moreover, these terms were also shown to be frequency independent in the exact theory such that the adiabatic approximation in this limit is in fact exact.
In view of these favorable properties of the ASCE kernel
the natural question arises whether this kernel can be used to correctly predict the excitation energies of dissociating molecules. Answering this question is the main aim of the present work.

To attack this problem, we developed a simplified one-dimensional model of a diatomic molecule having the main physical characteristics of a real three-dimensional hydrogen molecule and for which we can perform analytical and numerical calculations for arbitrary bond distance and interaction strength. In particular the KS orbitals and eigenvalues are known analytically, a feature that is very desirable as it provides an analytic expression for the KS gap upon dissociation. The model is used to benchmark the performance of the ASCE kernel as well as to discuss many features of the SCE formalism in the limit of large interactions.

The paper is organized as follows: In Sec.~\ref{sce_limit} we give a brief introduction to the main elements of SCE theory that we need.
In Sec.~\ref{sec2} we introduce the model system and discuss its properties.
In Sec.~\ref{sec4} we discuss
the ASCE kernel for our model density and obtain the excitation energies. In Sec.~\ref{sec5} we present our conclusions.

\section{Density-functional theory in the large interaction limit}
\label{sce_limit}

The main motivation of this work is to benchmark the recently proposed approximations for the exchange-correlation (xc) potential and xc-kernel based on the so-called theory of
strictly correlated electrons~\cite{Lani,Cort}. To provide a self-contained minimal background for the reader we briefly review some basic aspects of DFT. Our starting point is
the time-independent N-body Hamiltonian of a system which we write as~\cite{Barth:review}:
\begin{equation}
\hH_\lambda = \hT + \hV_\lambda + \lambda \hW
\label{Hlambda}
\end{equation}
where $\hT$ is the kinetic energy and  $\hW$ the two-body interaction, the strength of which is regulated by a real parameter $\lambda$. Finally, $\hV_\lambda$ represents
the external potential and is the sum of one-body potentials $v_\lambda (\br)$. The latter potential depends on the interaction strength $\lambda$ via the requirement that for
each value of $\lambda$ the same electronic density $n(\br)$ is obtained from the ground state of Eq.(\ref{Hlambda}). This makes $v_\lambda$ a functional of the density
via the Hohenberg-Kohn theorem \cite{HK} and we will therefore sometimes write $v_\lambda [n]$ to stress this fact when necessary.

Typically the Hamiltonian is given at $\lambda=1$ with a 
known external potential and the key many-body problem is to solve for its eigenstates. However, consideration of the full $\lambda$-dependence is useful in formal derivations in DFT and
is particularly relevant for the present work. An important limit is obtained by taking $\lambda=0$, in which case the system becomes non-interacting while retaining the
density of the interacting system. This system is denoted as the Kohn-Sham (KS) system and its external potential is commonly denoted by $v_s (\br)$. The ground state
of the KS system is a Slater determinant consisting of KS orbitals $\varphi_i$ satisfying
\begin{equation}
\left(  - \half \nabla^2 + v_s [n] (\br) \right) \varphi_i (\br,\sigma) = \epsilon_i \, \varphi_i (\br,\sigma)  
\label{KS_eqns}
\end{equation}
where $\sigma$ is a spin index. The KS equations are a device for obtaining the density of the interacting system by
solving one-particle equations. However, to make the procedure useful we need to make a connection to the interacting system
which we will take at a general interaction strength $\lambda$. To do this we define the Hxc potential as
\begin{equation}
v_{\Hxc}^{\lambda} [n] (\br) = v_s [n] (\br) - v_\lambda [n] (\br).
\end{equation}
A given approximation for this quantity allows us to obtain the density of the interacting system by
using the potential $v_{\textrm{KS}} [n, v_\lambda] = v_\lambda + v_{\Hxc}^{\lambda} [n]$ in Eq.(\ref{KS_eqns}) instead of $v_s [n]$ where
$v_\lambda$ is a given and known potential of the interacting system at interaction strenght $\lambda$  (which is commonly
taken to be $\lambda=1$ but we would like here to use a general interaction strength for the discussion below)~\cite{Ruggenthaler:2013}.
The central object of DFT is therefore the Hxc potential. This quantity in turn is given by the functional derivative of the Hxc-energy with respect to
the density $v_\Hxc^\lambda (\br)= \delta E_\Hxc^\lambda /  \delta n (\br)$. The Hxc-energy can be obtained from
\begin{equation}
E_{\text{Hxc}}^{\lambda}[n]=\int_{0}^{\lambda}d\lambda' W_{\lambda'}[n] 
\end{equation}
where we defined
\begin{equation}
W_\lambda [n]=\langle \Psi_\lambda [n]|\hat{W}|\Psi_\lambda [n]\rangle.
\end{equation}
where $\Psi_\lambda [n] $ is the ground-state of Hamiltonian Eq.(\ref{Hlambda}).  The quantity $W_\lambda$ has been studied in limiting cases.
For small values of $\lambda$ it is accessible via perturbation theory while in the limit of large values of $\lambda$
there is an asymptotic expansion that is derived from SCE theory.
This expansion has the form~\cite{GoriGiorgi:2009}
\begin{equation}
W_\lambda [n] = V_\textrm{SCE}[n]+\frac{V_\textrm{ZPE}[n]}{\sqrt{\lambda}} + O (\lambda^{-3/2} )
\label{wmu}
\end{equation}
where the leading term is the interaction energy of the strictly correlated electrons and the next term arises from
their zero-point energy (ZPE) in vibrations around their equilibrium positions.
Correspondingly the asymptotic expansion of the Hxc energy for large $\lambda$ is given by:
\begin{equation}
E_\Hxc^\lambda[n] = \lambda V_\textrm{SCE}[n]  + 2 \sqrt{\lambda} V_\textrm{ZPE}[n] + E_{2}[n]+ O( \lambda^{-1/2} ).
\label{elambda}
\end{equation}
as can be checked by differentiation with respect to $\lambda$ and comparison to Eq.(\ref{wmu}). This expression further introduces a density functional
$E_2 [n]$ the relevance of which will become clear later. 
The functional derivative with respect to the density gives an expansion of the Hxc-potential in powers of $\sqrt{\lambda}$ 
\begin{equation}
v_{\text{Hxc}}^{\lambda}(\br)=\lambda v_{\textrm{SCE}}(\br)+\sqrt{\lambda}v_{\textrm{ZPE}}(\br)+v_{2}(\br)+ O\left(\lambda^{-1/2}\right)
\label{PotentialExpansionSCE}
\end{equation}
which is valid for large value of $\lambda$. A very interesting point is that, at least for one-dimensional systems many-electron systems, the two leading terms are 
explicitly known functionals of the density and can be calculated explicitly in a rather simple way from so-called co-motions functions~\cite{Seidl1}. 
Before we discuss the applicability of this expansion let us further define the adiabatic Hxc kernel by
\begin{equation}
f_\Hxc^\lambda (\br,\br') = \frac{\delta v_\Hxc^\lambda (\br)}{\delta n(\br')}
\end{equation}
which according to Eq.(\ref{PotentialExpansionSCE}) has the expansion
\begin{equation}
f_\Hxc^\lambda (\br,\br') = \lambda \frac{\delta v_\textrm{SCE} (\br)}{\delta n(\br')} + O (\sqrt{\lambda})
\label{asce_kernel}
\end{equation}
The first term on the right hand side represents the so-called adiabatic SCE kernel $\lambda f^{\textrm{ASCE}}_\Hxc$ which has been studied in detail in Refs.\cite{Lani,Cort}
which we refer to for more details.
So far our discussion has been very general and, apart from the adiabatic approximation to the time-dependent kernel of TDDFT in Eq.(\ref{asce_kernel}), no
approximations have been used. 
The main question is, however, how reliable the asymptotic expansions in Eqs.(\ref{elambda}) and (\ref{PotentialExpansionSCE}) are for values
close to the physically relevant interaction strenght $\lambda=1$.
Since the expansion is asymptotic, retaining higher order terms typically worsen the approximation unless we increase the value of $\lambda$.
This means that for values of $\lambda$ close to one the best approximation may be obtained by only retaining the term $v_\textrm{SCE}$.
Indeed, it was pointed out in Ref.\cite{Malet} that in this interaction regime adding the ZPE contribution generally will give a worsening of the result.
It was found that at the lowest SCE level for a model one-dimensional diatomic molecule the bonding curve is correct at large separation but inaccurate
at equilibrium separation, while adding the ZPE contribution gives an overall worse result for the bonding curve. The asymptotic expansion can therefore
not been applied as such and consequently Ref.\cite{Malet} considers various amendments.
A similar conclusion was obtained from our previous work on the model system of a quantum ring~\cite{Cort} where we found the ZPE contribution to worsen
the results at smaller interaction strengths. This work was done for a homogeneous system in which we mainly studied the properties of the kernel
itself. In the present work we extend that work to an inhomogeneous model system in which again the kernel will be at the focus of attention.
The equations derived in the present section will be referenced in later sections.

\section{The molecular model}\label{sec2}
\subsection{Definition of the model}
For our description of the simplified molecular model we consider two electrons with spatial coordinates $x_1$ and $x_2$ both in the domain $\left[-\frac{L}{2},\frac{L}{2}\right]$ on a ring of length $L$.
The Hamiltonian of our system is given by
\begin{align}
\hat{H}_{\lambda} =&-\frac{1}{2}\left(\partial^2_{x_1}+\partial^2_{x_2}\right)+v_{\ssc{\lambda}}(x_1)+v_{\ssc{\lambda}}(x_2)\nonumber\\
&+\lambda\cos^2\left[\frac{\pi}{L}(x_1-x_2)\right]
\label{HamiltonianH2LambdaL}
\end{align}
where the first two terms are the kinetic energy of each electron, $v_{\lambda}$ is the one body external potential and $w (x) =\lambda \cos^2 (\pi x/L)$ is the electron-electron repulsion. 
We impose periodic boundary conditions such that the particles effectively move on a ring which
is commonly referred to as a quantum ring (QR) system \cite{Ruggenthaler:2013}.
The strength of the interaction $\lambda$ is a parameter which we will take to be positive. The interaction tends to keep particles on opposing parts of ring
and has a convenient form for numerical considerations. In accordance with Eq.(\ref{Hlambda}) the potential $v_\lambda$ is chosen in such a way that for each value of $\lambda$ the same ground state density is produced.
For our model it turns out to be useful to specify the external potential at $\lambda=0$ which corresponds to the KS-potential. In this way we can choose
the potential in such a way that we obtain an analytic solution for the KS orbitals. The potential at all other interaction strengths, including the physically relevant case $\lambda=1$, is
subsequently determined by the constraint that the density is the same for all values of $\lambda$ as we will discuss in more detail later. 

\subsection{The Kohn-Sham system}

The KS system is obtained from Eq.(\ref{HamiltonianH2LambdaL}) by taking $\lambda=0$ and we adopt the common notation of denoting the 
KS -potential by $v_s$, i.e. $v_s = v_{\lambda=0}$. In this limit the Hamiltonian of Eq.(\ref{HamiltonianH2LambdaL}),  which we now denote by $\hat{H}_s$, attains the form
\begin{equation}
\hat{H}_{s}  =-\frac{1}{2}\left(\partial_{x_1}^2+\partial_{x_2}^2\right)+v_{s}(x_1)+v_{s}(x_2) . \label{KShamiltonian}
\end{equation}
We now specify an explicit choice for $v_s$ which we take to be
\begin{equation}
v_{\ssc{s}}(x)=V_0\left[ 1 + \cos\left(\frac{4\pi x}{L}\right) \right]\label{KSpotential}
\end{equation}
where $V_0$ is a constant with units of energy. 
This potential has two minima located at $x_0=\pm L/4$ where $v_s (x_0) =0$ and is positive everywhere else. 
The ground-state density has two maxima at the potential minima and therefore represents a simple model of a diatomic molecule in which the atoms are
separated by a bond distance $L/2$.
\begin{figure}[ht]
\centering
\includegraphics[width=0.45\textwidth]{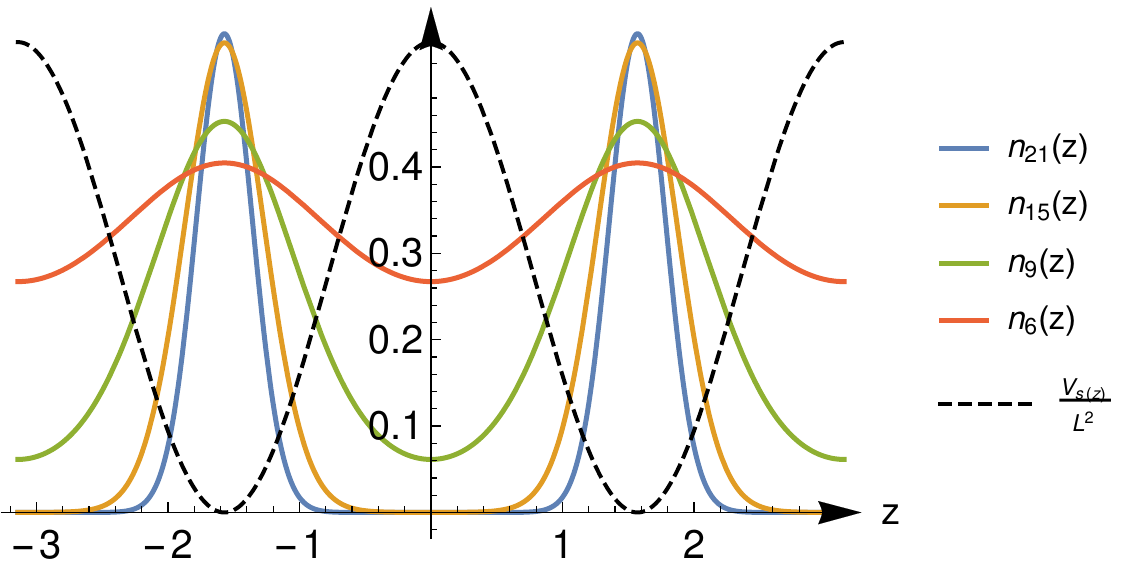}
\caption{ The ground-state density $n$ as a function of the dimensionless coordinate $z=2\pi x/L$. 
The densities for various $L$-values are denoted by $n_L$ in the plot.
The corresponding KS potential $v_s/L^2$ Eq.(\ref{KSpotential}) is plotted in arbitrary units for comparison and indicated by a dashed line.
For large $L$ we obtain two peaks of fixed width while for small $L$ the system becomes homogeneous.}\label{kspotentialdensity}
\end{figure}
We want to use this model to describe molecular dissociation and therefore vary the bond length $L$. While doing this we want to guarantee that the
width of each atomic density remains fixed upon separation, which can be achieved by requiring that the curvature of the potential at $x_0=\pm L/4$ is independent of $L$. This condition reads
\begin{align}
v''_{s}(x_0) =\left(\frac{4\pi}{L}\right)^2 V_0 = \alpha 
\label{ConditionWidth}
\end{align}
where $\alpha$ length independent which gives $V_0=\alpha\left(L/(4\pi)\right)^2$ for an arbitrary $\alpha$ (in this paper we will always take $\alpha=1$).
The KS orbitals of our system satisfy the eigenvalue equation
\begin{equation}
\left[- \frac{1}{2} \partial_x^2+ v_s (x) \right]\varphi_l^{\pm} (x)= \varepsilon_l^{\pm} \varphi_l^{\pm}(x) \label{KSequations}
\end{equation}
where we added a symmetry label $\pm$ for orbitals that are even or odd with respect to reflection in the origin, i.e. $\varphi_l^{\pm}(-x)=\pm \varphi_l^{\pm}(x)$.
These equations must be solved together with the boundary conditions $\varphi_l(-L/2)=\varphi_l(L/2)$ and the same for their derivatives. 
It is convenient to define the dimensionless coordinate $z=2\pi x/L$ and use the explicit form of the potential to rewrite Eq.(\ref{KSequations}) as
\begin{align}
\left[ - \partial_z^2+2\nu\cos(2z) \right]M^{\pm}_l(z)=a_l^{\pm} (\nu) M^{\pm}_l(z)\label{mathieuequation} 
\end{align}
where we have defined the following constants
\begin{align}
a_l^{\pm} (\nu) &=2\left(\frac{L}{2\pi}\right)^2(\varepsilon_l^{\pm}-V_0) \label{math_char}\\
\nu(L) &=\frac{\alpha}{4}\left(\frac{L}{2\pi}\right)^4  \label{nu} .
\end{align}
We recover the KS orbitals from $\varphi_l^\pm (x) = M_l^\pm (2 \pi x /L)$.
Equation (\ref{mathieuequation}) is the well-known Mathieu equation and its eigenfunctions and eigenvalues have been intensively studied~\cite{NIST}.
The functions $M_l^+$ and $M_l^-$ are commonly denoted as the Mathieu-cosine $C_l$ and the Mathieu-sine $S_l$ functions respectively,
while the values $a_l^\pm$ are called the Mathieu characteristic values.  The convention is that the label of the even states start at $l=0$ whereas
the labels of the odd states start at $l=1$.
The Mathieu functions satisfy $M_l^{\pm}(z+\pi)=(-1)^l M_l^{\pm}(z)$ and are therefore $2\pi$-periodic. They are commonly normalized as follows
\begin{equation}
\int_{-\pi}^{\pi}dz \, ( {M^{\pm}_l}(z) )^2 =\pi  \label{normalizationCondition}.
\end{equation}
Correspondingly the normalized (to one) KS orbitals are expressed in terms of Mathieu functions as
\begin{align}
\varphi_l^{+}(x)&=\sqrt{\frac{2}{L}}C_l\left(\frac{2\pi x}{L};\nu\right) \label{KScos}\\
\varphi_l^{-}(x)&=\sqrt{\frac{2}{L}}S_l\left(\frac{2\pi x}{L};\nu\right)
\label{KSorbitals} 
\end{align}
while the Kohn-Sham eigenenergies can be recovered from the Mathieu characteristic values 
by means of Eq.(\ref{math_char}).
In \Fig{kspotentialdensity} we plot the KS potential and the ground state density for different bond distances to illustrate the main features that we mentioned, in particular the fact that 
the width of the maxima becomes independent of the bond distance for large $L$.
Although we are not particularly interested in the case of very short bond distances we note that in the limit $L \rightarrow 0$ the parameter $\nu$ becomes  equal to zero and the ground state KS orbital is
given by the constant function $\varphi_0 (x) = 1/\sqrt{L}$ representing a system of constant density. We will not investigate this limit in detail; a homogenous QR at various interaction strength
has been studied in detail in Ref.\cite{Cort}.\\
In Fig.\ref{Fig2} we plot the ground state and the first few excited state KS orbitals. Of particular interest for our later discussion of the Hxc kernel is the lowest pair of bonding and anti-bonding states
represented by the pair of Mathieu functions $C_0$ and $S_1$. 
\begin{figure}[ht]
\centering
\includegraphics[width=0.45\textwidth]{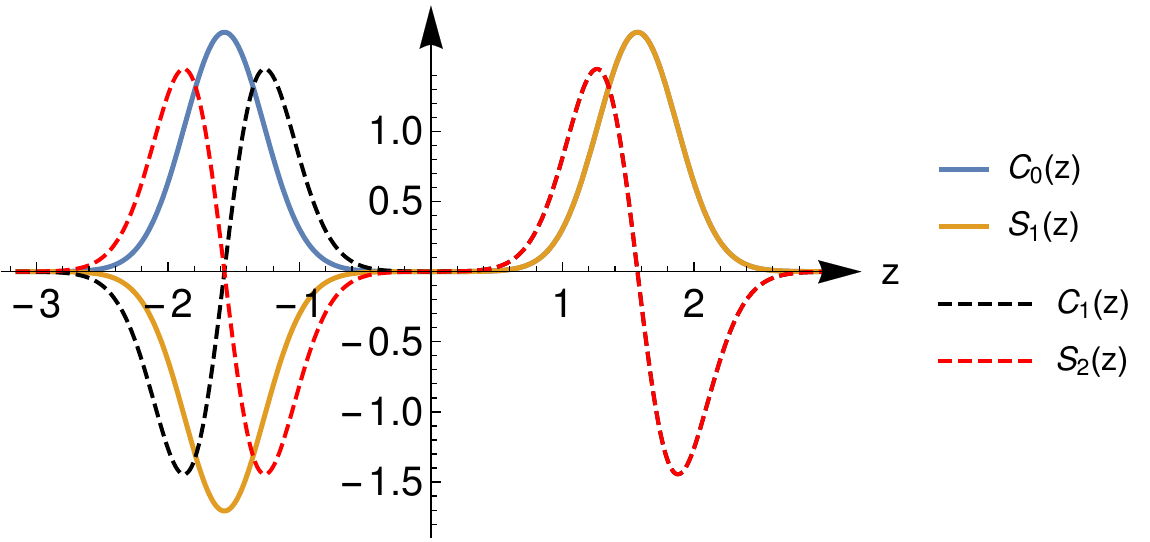}
\caption{Selected KS orbitals as a function of the dimensionless coordinate $z=2 \pi x/L \in [-\pi,\pi]$ plotted for the bond distance $L/2=10.5$. We display the ground state and first few excited states
corresponding to the bonding and anti-bonding orbital pairs
represented by the pair of Mathieu functions $C_0$ and $S_1$ as well as the pair $C_1$ and $S_2$. For this bond distance the bonding and anti-bonding orbitals
coincide for positive $z$.
 }
\label{Fig2}
\end{figure}
 The corresponding energy gap between the KS eigenvalues closes exponentially fast with increasing bond distance: 
\begin{align}
\varepsilon_1^- - \varepsilon_0^+ &= \frac{2 \pi^2}{L^2} (a_1^- (\nu)- a_0^+ (\nu) ) \\
& = \frac{32}{L^2} (2\pi)^{3/2} \nu^{3/4}e^{-4\sqrt{\nu}} \xrightarrow[L \to \infty]{} 0 
\end{align}
where we used the asymptotic expansion for the Mathieu characteristic value given in (\ref{appendixA}).
We remind the reader that $\nu$ is an increasing function of $L$ given by Eq.(\ref{nu}).
The density in the bond midpoint has a similar exponential decay (see Eq.(\ref{cooo}) in Appendix \ref{appendixA}) 
given by
\begin{equation}
n(0) = \frac{16}{L} \left( 2 \pi \right)^{\half}  \nu^{1/4} e^{-4 \sqrt{\nu} } \quad (L \rightarrow \infty)
\label{eq_n0}
\end{equation}
The knowledge of this precise behavior of the KS gap as well as the density in the bond midpoint 
will facilitate considerably the calculation of the
excitation energy from the ASCE kernel in Sec.\ref{sec4}.

\subsection{Exact solution of the model}\label{sec3}

After having considered the model in the KS limit we will now consider the case of finite interaction strength $\lambda$. The potential $v_\lambda$ in Eq.(\ref{HamiltonianH2LambdaL}) can not be obtained
analytically except in some limiting cases that we will discuss below. We therefore obtain $v_\lambda$ directly from the constraint that the density is independent of $\lambda$ 
using the numerical algorithm outlined in Ref.\cite{Nielsen:EPJB2018}. In our case the density is given by the ground state KS orbital from Eq.(\ref{KScos}) to be
\begin{equation}
n(x) = 2 |\varphi_0^+ (x)|^2 = \frac{4}{L} \, C_0^2 (\frac{2\pi x}{L};\nu)
\end{equation}
for all $\lambda$ where we remind the reader that $\nu$ depends on $L$ via Eq.(\ref{nu}). Therefore for a given value of $L$ we have the numerical task to find $v_\lambda$ for a range of interaction strengths of interest.
The ground state is a spin singlet state and consequently we will mostly be interested in the singlet excited states. The singlet wave function 
has the structure
\begin{equation}
\nonumber
\Psi(x_1\sigma_1,x_2\sigma_2) =\psi (x_1,x_2) \frac{1}{\sqrt{2}}\left(\delta_{\sigma_1\uparrow}\delta_{\sigma_2\downarrow}-\delta_{\sigma_1\downarrow}\delta_{\sigma_2\uparrow}\right)
\end{equation}
where $\sigma_i$ for $i=1,2$ are spin variables and where the spatial part of the wave function is symmetric $\psi (x_1,x_2)=\psi (x_2,x_1)$ to ensure anti-symmetry of the full space-spin wave function.
To obtain deeper insight in the results we will also derive analytic results in the regime of large bond distance $L/2$ for fixed interaction strength $\lambda$ which is the common molecular dissociation regime 
and the complementary regime of large interaction strength $\lambda$ for fixed bond distance $L/2$ which is the SCE regime. 
We will start in the next subsection with the first regime.

\subsubsection{Large bond distance for fixed interaction strength}

We first consider the regime of large bond distance $L/2$ at fixed values of $\lambda$.
In this regime the molecule is typically dissociated in two one-electron atoms (unless the interacting strength $\lambda$ is very small such that there
are contributions from the ionic states with two or zero electrons on each atom).
For a one-electron atom the KS potential is equal to the true external potential and
therefore we have $v_\lambda (x) = v_s (x)$ for $x$ in the neighbourhood of each atom at large separation.
The ground state atomic orbitals $A(x)$ and $B(x)$ on atoms $A$ and $B$ are localized around $x=\pm L/4$ and
can be expressed in terms of the first bonding and anti-bonding molecular KS orbitals as $A(x) = (\varphi_0^+ (x) + \varphi_1^- (x))/\sqrt{2}$
and $B(x) = (\varphi_0^+ (x) - \varphi_1^- (x))/\sqrt{2}$ (see for example Fig.\ref{Fig2}).
The exact ground-state (GS) wave function for the large bond distance limit is the well known Heitler-London (HL)  wave function 
\begin{align}
\Psi_{\lambda}^{\textrm{GS}}(x_1,x_2) &= \frac{1}{\sqrt{2}} \left[ A (x_1) B (x_2)+ B (x_1) A(x_2) \right]\nonumber \\
&=\frac{1}{\sqrt{2}} \left[\varphi^+_0(x_1)\varphi^+_0 (x_2)-\varphi^-_1(x_1)\varphi^-_1(x_2) \right]  \label{HLGroundState}
\end{align}
The ground state energy is given by
\begin{equation}
E_\lambda^\textrm{GS} = 2 \varepsilon_0^+ = \sqrt{\alpha} - \frac{\pi^2}{L^2} + O ( L^{-4})  \quad (L \rightarrow \infty)
\label{eq:gs}
\end{equation}
where we used that $\varepsilon_0^+=\varepsilon_1^-$ in the large $L$ limit and the asymptotic expansion
of the Mathieu characteristic values in Appendix \ref{appendixA}. This result is easy to understand. Since at the atomic positions
$x_0=\pm L/4$ we have that $v_s^{''} (x_0)=\alpha$ the potential around each atom is given by $v_s (x) = \alpha (x-x_0)^2 /2$ which
corresponds to a harmonic well with harmonic frequency $\sqrt{\alpha}$. Each atomic oscillator has ground state energy $\sqrt{\alpha}/2$ 
thereby adding up to the molecular ground state energy $\sqrt{\alpha}$.

Let us now consider the first excited state which in the large $L$ limit is given by
\begin{align}
\Psi_{\lambda}^{(1)}(x_1,x_2)=&\frac{1}{2}\left[\varphi^+_1(x_1) \varphi^-_1(x_2)+\varphi^-_1(x_1)\varphi^+_1(x_2)\right.\nonumber\\
&\left.-\varphi^+_0(x_1)\varphi^-_2(x_2)-\varphi^-_2(x_1)\varphi^+_0(x_2)\right] .
\label{HLexcitedk}
\end{align}
The orbitals used in this expression are displayed in Fig.\ref{Fig2}.
For large $L$ the states $\varphi_0^+$ and $\varphi_1^-$ become degenerate  and the same is true for the states
$\varphi_1^+$ and $\varphi_2^-$.  These orbitals can be used to construct localized ground and excited state atomic orbitals
from the combinations $\varphi_0^+ \pm \varphi_1^-$ and $\varphi_1^+ \pm \varphi_2^-$ if desired.
The energy of the two-particle state of Eq.(\ref{HLexcitedk}) is given by
\begin{equation}
E_\lambda^{(1)} = \varepsilon_0^+ + \varepsilon_1^+ =  2 \sqrt{\alpha} - \frac{3\pi^2}{L^2} + O ( L^{-4})  \quad (L \rightarrow \infty)
\label{eq:excited}
\end{equation}
Again it is straightforward to interpret the energy. The system is a superposition of two states in which one atom is a 
ground state oscillator with energy $\sqrt{\alpha}/2$ and the other one a first excited oscillator with energy $3 \sqrt{\alpha}/2$
giving a total molecular energy of $2 \sqrt{\alpha}$.

\begin{figure}[ht]
\includegraphics[width=0.45\textwidth]{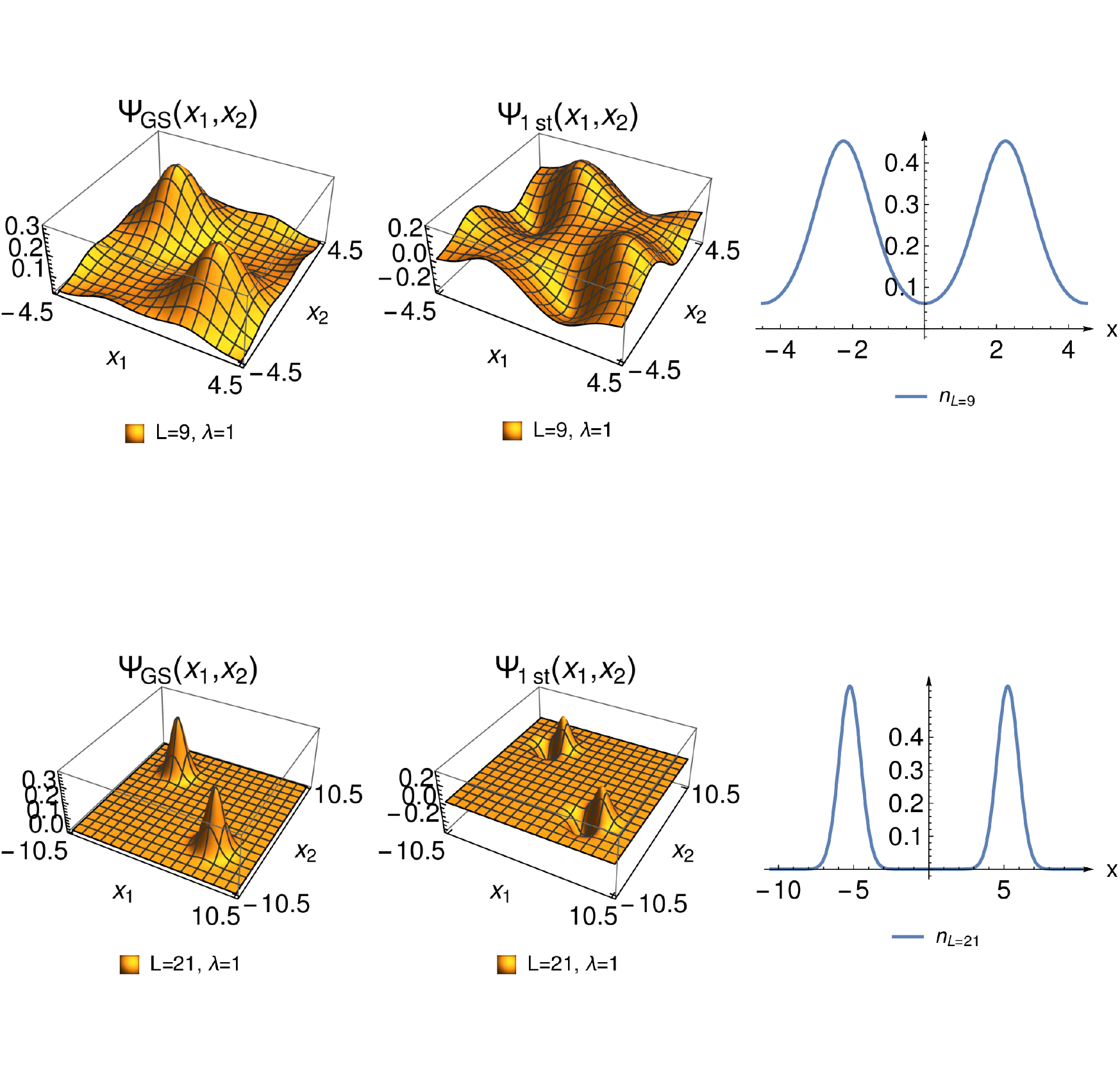}
\caption{The ground and first excited state wave functions for interaction strength $\lambda=1$ plotted for $L=9$ and $L=21$.
The rightmost panels display the corresponding ground state densities.}
\label{Fig3}
\end{figure}

To judge the accuracy of these limiting wave functions we plot the exact $\psi_\lambda$ for  $\lambda=1$ and $L=9$ and $21$ (corresponding to bond lengths $4.5$ and $10.5$) 
in Fig. \ref{Fig3}. We see that for $L=21$ the wave functions Eq.( \ref{HLGroundState}) and
(\ref{HLexcitedk}) are a good approximation to the true wave functions (as we also checked numerically).  At $L=9$ the system still has a considerable density at the bond midpoint
and the HL-type wave functions are a less good approximation.

\begin{figure}[ht]
\includegraphics[width=0.35\textwidth]{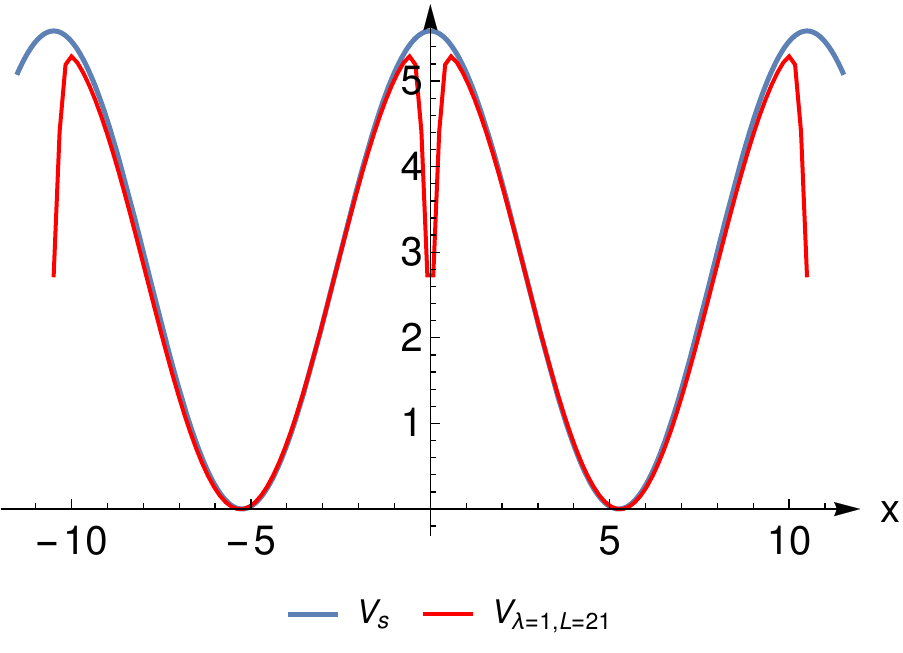}
\caption{The potential $v_\lambda$ for $\lambda=1$ for $L=21$ compared to $v_s$. }
\label{Fig4}
\end{figure}

Finally we compare in Fig.\ref{Fig4} the exact external potential $v_\lambda$ to $v_s$. We see that around the atoms both potentials agree but that around the bond midpoint there
is a considerable deviation. This amounts to a peak in the Hxc-potential $v_\Hxc^\lambda = v_s - v_\lambda$ at the bond midpoint. This is a well-known feature of the Hxc-potential 
\cite{BuijseBaerends} and is related to the so-called left-right correlation in the system. We refer to the cited reference for a more in-depth discussion.
%


\subsubsection{Large interaction strength at fixed bond distance}

\begin{figure}[ht]
\includegraphics[width=0.45\textwidth]{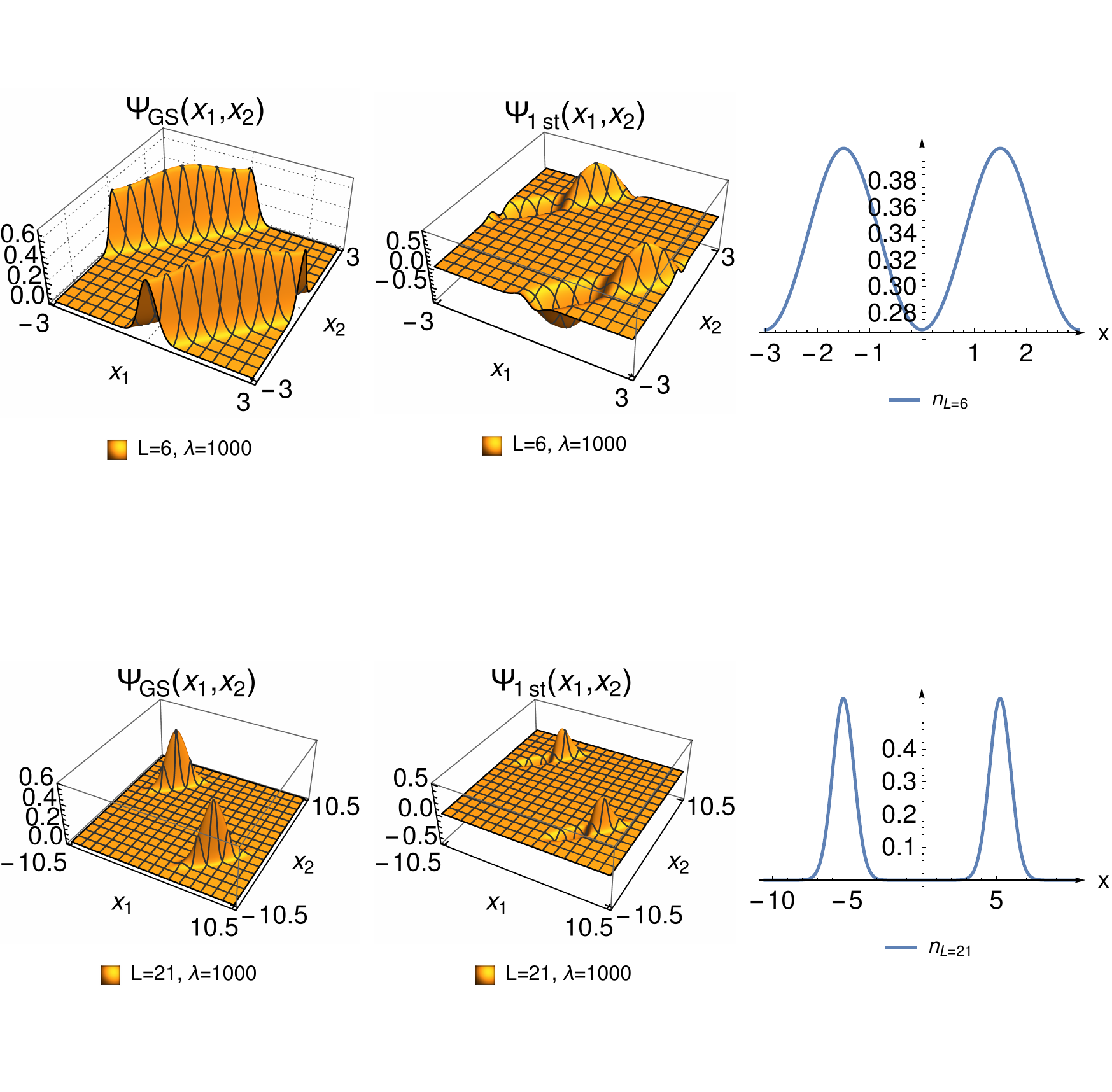}
\caption{The ground and excited state wave functions at large interaction strength $\lambda=1000$ for the bond distances $L/2=3$ and $L/2=10.5$. We
note that the wave function localizes in narrow strips along the lines $|x_1-x_2|=L/2$. The right most panels display the corresponding ground state densities.}
\label{Fig5}
\end{figure}

\begin{figure}[ht]
\includegraphics[width=0.35\textwidth]{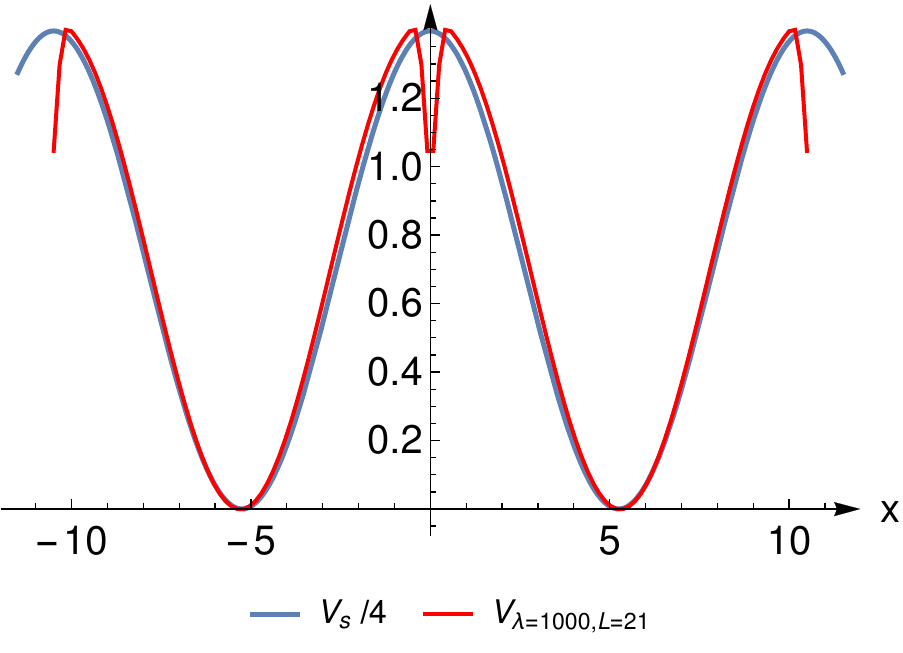}
\caption{The potential $v_\lambda$ for $\lambda=1000$ for $L=21$ compared to $v_s / 4$.}
\label{Fig6}
\end{figure}

We now turn our attention to the complementary regime of larger interaction strength $\lambda$ for fixed bond distance. This is the regime in which SCE become exact.
From our numerical work we find that in this limit the two-particle wave function localizes in a region where $|x_1 -x_2| \approx L/2$ as displayed
in \Fig{Fig5}. 
 This is in accordance with SCE theory which
tells that in the very strong interaction limit the position of a single electron determines the positions of the remaining electrons uniquely.
 For this reason it is convenient to introduce the center of mass $R=(x_1+x_2)/2$ and relative coordinate $r=x_1-x_2$, where $R \in [-L/2,L/2]$ and $r \in [-L,L]$. 
 The Hamiltonian (\ref{HamiltonianH2LambdaL}) 
in the new coordinates attains the form
\begin{align}
\hat{H}_{\lambda} =&-\frac{1}{4}\partial_R^2-\partial_r^2+v_{\ssc{\lambda}}\left(R+\frac{r}{2}\right)+v_{\ssc{\lambda}}\left(R-\frac{r}{2}\right)\nonumber\\
&+\lambda\cos^2\left(\frac{\pi r}{L}\right)\label{HamiltonianH2LambdaLTrans}
\end{align}
We want to give an explicit approximate expression of the hamiltonian (\ref{HamiltonianH2LambdaL}) for the limit $\lambda\to\infty$ for any fixed bond distance $L/2$. 
Since the wave function is localized around the lines $r=\pm L/2$ it is natural to expand the external potential $v_{\lambda}$ around these values.
For example, for $r=L/2$ we have to second order
\begin{align} 
    v_{\lambda}\left(R+\frac{r}{2}\right)+&v_{\lambda}\left(R-\frac{r}{2}\right) \nonumber\\
=& \bar{v}_{\lambda} (R)+ \beta_\lambda (R) \left(r - \frac{L}{2}\right)^2\label{ExpansionPotentialLargeLambda}                                                                                                                                                                                                                                                                                                    \end{align}
where we defined
\begin{align}
\bar{v}_\lambda (R) &= 2 \, v_{\lambda}\left(R+ \frac{L}{4}\right) \\
\beta_\lambda (R) &=  \left. \frac{\partial^2 v_\lambda (R\pm r/2)}{\partial r^2} \right|_{r=L/2}
\end{align}
with an essentially identical result for the expansion around $r=-L/2$,
and where we used the property $v_\lambda (x) = v_\lambda (x+L/2)$ in the definitions of $\bar{v}_\lambda$ and $\beta_\lambda$ and in the cancellation of the linear term.
With the expansion of Eq.(\ref{ExpansionPotentialLargeLambda}) the Hamiltonian becomes
\begin{align}
\hat{H}_{\lambda} =&-\frac{1}{4}\partial_R^2 - \partial_r^2+ \bar{v}_{\lambda} (R)+ \beta_\lambda (R) (r- \frac{L}{2})^2 \nonumber\\
&+\lambda\cos^2\left(\frac{\pi r}{L}\right)\label{HamiltonianH2LambdaLTrans}
\end{align}
with a similar expansion around $r=-L/2$. We see that this Hamiltonian becomes separable when we neglect the term $\beta_\lambda$.
However, the two-body interaction has form $w (r) = \lambda (\pi/L)^2 (r- L/2)^2$ around $r=L/2$ and the question is therefore whether we can neglect
$\beta_\lambda$ compared to $\lambda (\pi/L)^2$. From our calculation we find that $v_\lambda$ and therefore also $\beta_\lambda$ converges to
a finite value for large $\lambda$. Therefore for fixed $L$ and large enough $\lambda$ we can neglect $\beta_\lambda$ and the system becomes
approximately separable.
If we write the wave function in this limit as $\Psi_\lambda (r,R) = \chi_\lambda (r)  \varphi_\lambda (R)$ then its factors are determined from the equations
\begin{align}
\left(  -\frac{1}{4}\partial_R^2 + 2\, {v}_{\lambda} (R + \frac{L}{4}) \right) \varphi_\lambda (R) = \epsilon \, \varphi_\lambda (R) 
\label{ham1}\\
\left( - \partial_r^2 + \lambda\cos^2\left(\frac{\pi r}{L}\right) \right) \chi_\lambda (r) = \tilde{\epsilon} \, \chi_\lambda (r)
\label{ham2}
\end{align}
These equations determine all the eigenstates in the large $\lambda$ limit. Let us, however, focus on the ground state
and take $\chi_\lambda$ and $\varphi_\lambda$ to be ground states of their corresponding Hamiltonians.
The ground state density is then obtained from
\begin{equation}
n(x_1) = 2 \int_{-L/2}^{L/2} dx_1 |\varphi_\lambda (\frac{x_1+x_2}{2})|^2 |\chi_\lambda (x_1-x_2)|^2
\label{int_dens}
\end{equation}
The function $| \chi_\lambda (r)|^2$ becomes very narrowly peaked around $r=\pm L/2$ as $\lambda$ becomes very large.
We can therefore normalize it such that for the limit that $\lambda \rightarrow \infty$
\begin{equation}
|\chi_\lambda (r) |^2  \rightarrow    \delta (r-\frac{L}{2}) + \delta (r+\frac{L}{2} )
\end{equation}
from which we obtain, using Eq.(\ref{int_dens}), that for large interaction strength
\begin{align}
n (x) &= 2 \left[   |\varphi_\lambda (x + \frac{L}{4})|^2   +   |\varphi_\lambda (x - \frac{L}{4})|^2 \right] \nonumber \\
&= 4 \, |\varphi_\lambda (x - \frac{L}{4})|^2
\end{align}
The ground state density is also given by $n(x) = 2 |\varphi_0^+ (x)|^2$ in which $\varphi_0^+ (x)$  solves Eq.(\ref{KSequations}).
Comparison of this equation to Eq.(\ref{ham1}) then immediately yields that
\begin{equation}
v_\lambda (x) = \frac{v_s (x)}{4} = \frac{V_0}{4} \left[ 1 + \cos\left(\frac{4\pi x}{L}\right) \right]
\end{equation}
and $\varphi_0^+ (x) = \sqrt{2} \, \varphi_\lambda (x-L/4)$. From our derivation we therefore deduce that in our system
\begin{equation}
\lim_{\lambda \rightarrow \infty} v_\lambda (x) = \frac{v_s (x)}{4} 
\end{equation}
A comparison with the general Eq.(\ref{PotentialExpansionSCE}) from SCE theory shows that in our case 
$v_\textrm{SCE}$ and $v_\textrm{ZPE}$ are zero and that $v_{2} (x) = v_s- v_\lambda = 3v_s (x)/4$. The fact that 
$v_\textrm{SCE}$ and $v_\textrm{ZPE}$ vanish can also be directly derived from SCE theory and is a consequence of
the symmetry of our system.
In Fig. \ref{Fig6} we compare $v_\lambda$ to $v_s/4$ for various large values of $\lambda$ and note a good agreement between them
with the exception of some deviations around the bond midpoint. This discrepancy becomes smaller for higher values of $\lambda$.

Let us now consider the energies of the system. The eigenenergies of the two-particle state are given by 
$E=\epsilon + \tilde{\epsilon}$ where $\epsilon$ and $\tilde{\epsilon}$ are the eigenvalues of Hamiltonians in
Eq.(\ref{ham1}) and  Eq.(\ref{ham2}). From the fact that $v_\lambda = v_s/4$ in Eq.(\ref{ham1}) we see that
the eigenvalues $\epsilon$ are half of the KS eigenvalues of Eq.(\ref{KSequations}). These eigenvalues
correspond to an excitation which only involves a change of the center-of-mass wave function without changing
the relative wave function. The eigenvalues $\tilde{\epsilon}$ are calculated from Eq.(\ref{ham2}).
The transformation $z=\pi r/L$ transforms this Hamiltonian to
\begin{equation}
\left[ - \partial_z^2 + 2 q  \cos (2z) \right] M (z) = a (q) M (z)
\label{Mathieu_again}
\end{equation}
where
\begin{align}
q &= \lambda \left( \frac{L}{2 \pi}\right)^2  \\
a(q) &= \frac{L^2}{\pi^2} \tilde{\epsilon} - 2q
\end{align}
Eq.(\ref{Mathieu_again}) is again the Mathieu equation with this time a parameter $q$ that depends on the interaction strength.
The eigenvalues in the limit of large interactions have the form
\begin{equation}
\tilde{\epsilon}_l = \frac{\pi^2}{L^2} (2q+ a_l^{+} (q)) = (l+\half) \frac{2\pi}{L} \sqrt{\lambda} \quad (\lambda \rightarrow \infty) 
\label{zpe_osc}
\end{equation}
which is a harmonic spectrum with harmonic frequency $\omega_\lambda = 2 \pi \sqrt{\lambda}/L$. These excitations of
involve a change of the relative wave function and represent the zero point vibrations of the strictly correlated electrons of 
SCE theory. The lowest  excitation energy for this mode is therefore $\omega_\lambda$.
This will be relevant of our discussion of the excitation energy obtained from the ASCE kernel.

\section{The adiabatic SCE kernel}
\label{sec4}

\subsection{Definition and properties}

We have in studied in detail the excitation properties of our model system in two different regimes.
We will now investigate the adiabatic SCE kernel. As was discussed
below Eq.(\ref{asce_kernel}) the ASCE kernel is defined as
\begin{equation}
f^\textrm{ASCE} (x,x') = \frac{\delta v_\textrm{SCE} (x)}{\delta n (x')}
\end{equation}
The SCE potential vanishes for our system, but its functional derivative does not. As was discussed in detail in Refs.\cite{Lani,Cort} it is explicitly given by
the expression
\begin{align}
f^{\textrm{ASCE}}(x,x') &=\int\limits_{-L/2}^{x} dy\ \frac{w''(y-f(y))}{n(f(y))} \\
& \times \left[\theta(y-x')-\theta(f(y)-x')\right]
\label{HxcKernel}
\end{align}
where $\theta$ is the usual Heaviside function and $w(x)$ the two body interaction.
The function $f(x)$ is the so-called co-motion function which specifies the position of
another electron given the position of a reference electron. For our system the
co-motion function attains the simple form
\begin{align}
f(x)=\left\{
\begin{array}{ccc}
x-\frac{L}{2}\ \text{if}\ x>0\\ 
x+\frac{L}{2}\ \text{if}\ x \leq 0
\end{array}
\right.\label{ComotionRing}
\end{align}
\begin{figure}[ht]
\centering
\includegraphics[width=0.5\textwidth]{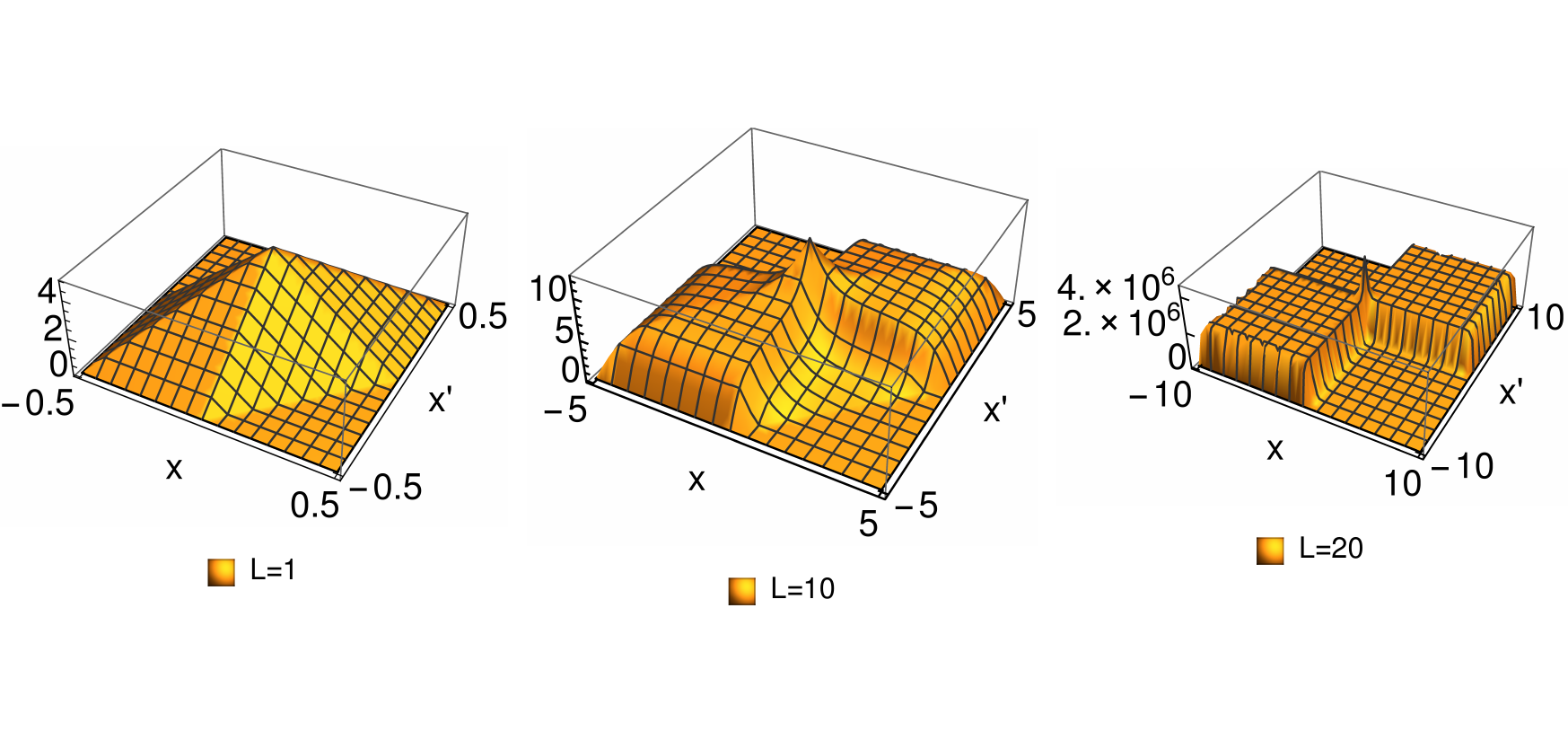}
\caption{The ASCE kernel for $L=1,10,20$. We see that with growing $L$ plateaux develop the heights of which grow exponentially with $L$.} 
\label{Fig8}
\end{figure}
If we define the function $\mathcal{P}(x)$ to be
\begin{equation}
\mathcal{P}(x) = \int\limits_{-L/2}^{x}dy\ \frac{w''(y-f(y))}{n(f(y))} 
\end{equation}
then the integrand contains $d \mathcal{P}/dx$ and we can obtain $f^\textrm{ASCE}$ by
partial integration while usefully manipulating the results using the fact that
$\mathcal{P} (x) - \mathcal{P}(0)$ is an odd function.
In the quadrant $x,x' >0$ we obtain
\begin{equation}
f^\textrm{ASCE} (x,x') = \mathcal{P} (-x) \theta (x-x') + \mathcal{P} (-x') \theta (x'-x)
\end{equation}
while in the quadrant $x<0,x'>0$ we have
\begin{equation}
f^\textrm{ASCE} (x,x') = \left[\mathcal{P}(x)-\mathcal{P}(x') + \mathcal{P}(0)\right]\theta (f(x)-x')
\end{equation}
The function in the remaining quadrants is determined from the symmetry $f^\textrm{ASCE} (x,x') = f^\textrm{ASCE} (-x,-x')$.
For our system the function $\mathcal{P}(x)$ can be written more explicitly as:
\begin{align}
\mP(x)&=w''(\frac{L}{2}) \int_{-L/2}^{x}  \frac{dy}{n(y)}  
\label{px_func}
\end{align}
where for our two-body potential $w'' (L/2) = 2 \pi^2/L^2$.
In the Appendix \ref{AppendixB} we show that
\begin{equation}
\lim_{L \rightarrow \infty} \mathcal{P} (x) = \mathcal{P} (0) \left[  \half + \theta (x) \right]
\end{equation}
for $x \neq 0$.
This equation implies that for large values of $L$ the kernel assumes the form
\begin{equation}
f^{\text{ASCE}}(x,x')= \half \mathcal{P}(0)\left[\theta(x)\theta(x')+\theta(-x)\theta(-x')\right]
\label{HxcKernelLargeL}
\end{equation}
for $x,x' \neq 0$.
The function exhibits plateaux of height $\mathcal{P} (0)$ in the quadrants in which both coordinates have the same sign
and is zero otherwise. In the Appendix \ref{AppendixB} we show that this height grows exponentially fast with $L$
according to
\begin{equation}
\mathcal{P} (0) = w''(\frac{L}{2})  \frac{L^2}{16 (2 \pi)^{3/2} \nu^{3/4}}e^{4\sqrt{\nu}} \quad (L \rightarrow \infty)
\label{P_limit}
\end{equation}
(we remind the reader that $\nu$ depends on $L$ according to Eq.(\ref{nu})).
With these results we are ready to calculate excitation energies from the ASCE kernel.

\subsubsection{Lowest excitation energy}

We now address the issue of calculation the excitation energy of the system.
To make our point it is sufficient to restrict ourselves to the so-called small matrix approximation \cite{Ullrich:book} in which
the singlet excitation energy $\Omega$ from an occupied state $i$ to an unoccupied state $a$ is given by
\begin{equation}
\Omega^2 = \omega_{ia}^2 + 4 \omega_{ia} K_{ia,ia}
\end{equation}
where $\omega_{ia} =\epsilon_a -\epsilon_i$ is the difference in KS energies.
and 
\begin{equation}
K_{ia,ia} = \int dx dx' \Phi_{ia} (x) f_\Hxc (x,x') \Phi_{ia} (x')
\end{equation}
where $\Phi_{ia} (x) = \varphi_i (x) \varphi_a (x)$ is an excitation function (in which we take the orbitals
to be real for simplicity) and $f_\Hxc$ the Hxc kernel which we took in an adiabatic approximation
relevant to the discussion below.
In our particular case we consider the excitation from the
lowest KS orbital $\varphi_0^+$ to $\varphi_1^-$. For easy of notation and to be in accordance with
adopted language we denote the orbitals by the gerade and ungerade sigma orbitals $\sigma_g (x)$ and $\sigma_u (x)$
and their eigenvalues by $\epsilon_g$ and $\epsilon_u$.
We know that in the dissociation limit the KS gap $\omega_{gu}$ vanishes . The excitation energy is therefore given by
\begin{equation}
\Omega^2 = \lim_{L \rightarrow \infty} 4 \omega_{gu} K_{gu,gu}
\label{OmegaLimit}
\end{equation}
In the ALDA this expression vanishes as the kernel can not compensate for the decay of the KS gap.
However, as we will show now, the ASCE kernel (we remind the reader of Eq.(\ref{asce_kernel}) ) will lead to a finite contribution.
The matrix element in the large separation limit is readily calculated from Eq.(\ref{HxcKernelLargeL}) to be
\begin{equation}
K_{gu,gu} = \frac{\lambda}{4} \mP (0)
\label{matrix_element}
\end{equation}
where we used the symmetry and normalization of the KS orbitals.
If we use this in Eq.(\ref{OmegaLimit}) we find that in the large $L$ limit
\begin{equation}
\Omega^2 = \lambda  (\epsilon_u -\epsilon_g)  \mathcal{P} (0) = 2 \lambda w'' (\frac{L}{2}) \quad (L \rightarrow \infty)
\label{eq:limit}
\end{equation}
For our system we have $w'' (L/2) = 2 \pi^2 /L^2$ and we obtain $\Omega = 2 \pi \sqrt{\lambda}/L$
which is exactly the harmonic frequency of the zero point oscillation of Eq.(\ref{zpe_osc}). We therefore
deduce that the excitations that we recover from the ASCE kernel are exactly the ones that correspond to
the zero point oscillations. With hindsight this may not be surprising as, after all, the zero point oscillations
represent an always present set of excitations in SCE theory. Note that in the derivation of Eq.(\ref{eq:limit}) it is important to
consider a fixed but arbitrary large $L$ and then take the limit $\lambda \rightarrow \infty$, i.e. the standard SCE regime, 
and not the other way around otherwise $\Omega = 2 \pi \sqrt{\lambda}/L \rightarrow 0$.

\subsection{The ASCE kernel in the conventional molecular dissociation regime}

In the previous subsection we found that in the limit that the interaction strength $\lambda$ becomes
very large at fixed bond distance $L/2$ the lowest excitation energy is that of the lowest zero point oscillation
of the strictly correlated electrons, and in that regime the ASCE kernel gives an exact result.
Let us now see how the ASCE kernel performs in the opposite regime in which the bond distance becomes
large at fixed interaction strength, in particular for the chemically relevant case of interaction strength $\lambda=1$.
This is the conventional dissociation regime as commonly studied in bond breaking in chemistry.
Note that we now apply the ASCE kernel outside its formal range of applicability and therefore the approximation becomes
uncontrolled. The consideration is nevertheless illuminating as it illustrates the reasons for the breakdown of
the approximation.
For $\lambda=1$ the matrix element Eq. (\ref{matrix_element}) of the
ASCE kernel is given by $\mP (0)/4$ and we have for the lowest excitation energy
\begin{equation}
 \Omega^{\textrm{ASCE}} = (2  w'' (\frac{L}{2}))^\half  \quad (L \rightarrow \infty)
\label{excitation_ASCE}
\end{equation}
Let us compare this to the exact excitation energy
\begin{equation}
\Omega^{\textrm{exact}} = \sqrt{\alpha} \quad (L \rightarrow \infty)
\label{eq:exact}
\end{equation}
as follows directly from Eqs.(\ref{eq:gs}) and (\ref{eq:excited}). We remind the reader
that the parameter $\alpha$ (see Eq.(\ref{ConditionWidth})) is given by the curvature of
the external potential at its minima (as $v_s$ becomes the true external potential around the atoms
in the dissociation limit). Since upon dissociation the separate atoms become independent single particle oscillators, Eq.(\ref{eq:exact}) is
a natural result. If we consider the ASCE approximation, on the other hand, we see that according to Eq.(\ref{excitation_ASCE}) the lowest
excitation energy is determined solely by the curvature of the interaction potential $w'' (L/2)$.
This is because, by using the ASCE kernel, we pretend that the separated atoms still behave as strictly correlated electrons
with an excitation energy determined by the zero point oscillations. This is the wrong physical
picture in this regime and therefore the ASCE approximation fails to describe the right physics.
In fact, in our system $w'' (L/2) = 2 \pi^2 /L^2 \rightarrow 0$ for $L \rightarrow \infty$ and
therefore the ASCE excitation energy becomes zero in the dissociation limit. For
other forms of the two-body interaction this may not be the case but this does not change our conclusion
regarding the physical picture.
The ASCE approximation is therefore not an improvement over the ALDA in the dissociation regime.
Both approximations attain the wrong dissociation limit; in the case of the ALDA the excitation
energy becomes zero whereas in the case of the ASCE approximation the excitation energy is determined by
the two-body interaction potential rather than by the external potential of the separated atoms.
This result is not surprising as we have used the ASCE kernel outside its regime of
applicability. The ASCE kernel is therefore not of use if
one is interested in regime of large bond length at intermediate interaction strength
which is the relevant case for bond breaking in most common chemical applications.
To correct these problems within the present formalism a natural way to proceed would be include ZPE and
higher order kernels in the expansion of the Hxc kernel as was done in Ref.\cite{Cort}.
However, that work showed that the extra terms lead to worse approximation than just the ASCE approximation
for low interaction strengths, as is typical for an asymptotic expansion.
The description of the conventional dissociation regime using density-functional methods therefore remains a
challenging task.

\section{Conclusions}
\label{sec5}

In this work we studied the properties of an approximate adiabatic Hxc kernel based on the theory of strictly correlated electrons. To benchmark this approximation we 
studied a numerically and analytically solvable system which is able to simulate the main features of a dissociating molecule. 
We studied in detail the two-particle eigenstates in various limits and calculated the excitation spectrum in the limit of large interaction strength.
The ASCE kernel was shown to reproduce the so-called zero-point oscillation part of the spectrum. The attainment of this exact result  shows that the ASCE
kernel becomes exact in the this regime as we also concluded from earlier work~\cite{Cort}. However, most current interest in molecular dissociation in chemistry
is devoted to the complementary regime of large bond distance at intermediate interaction strength. In this regime the ASCE kernel is not suitable for obtaining
the excitation spectrum. We conclude that the description of molecular dissociation based on functionals founded on SCE theory remains a challenge for
the future.

\appendix
\section*{Appendices}
\section{Properties of Mathieu functions}
\label{appendixA}

In this Appendix we describe a few useful properties the Mathieu functions and their characteristic values that we use in the main text.
Many properties of these functions can be found in~\cite{NIST}.
The Mathieu characteristic values have the following expansion for large $q$ (where $q$ is the parameter in the Mathieu equation)
\begin{align}
a_l^{+} (q) , a_{l+1}^- (q)&=-2 q+2 (2l+1) \sqrt{q}-\frac{1}{4}\left(2 l^2 + 2l+1\right) \nonumber \\
&+ \frac{(2l+1)}{128 \sqrt{q}}\left((2l+1)^2+3\right) + O{q^{-1}} \label{eq:excitationEnergies}
\end{align}
The difference $a_{l+1}^- (q)-a_l^+ (q)$ is exponentially small in the large $q$ limit~\cite{NIST}
\begin{align}
& a_{l+1}^-(q)-a_l^+(q)= \frac{2^{4l+5}}{l!} \left( \frac{2}{\pi}\right)^{\half}
q^{\frac{l}{2}+\frac{3}{4}}\ e^{-4\sqrt{q}} \nonumber \\
& \times\left[1-\frac{6l^2+14l+7}{32 \sqrt{q}}+O\left(q^{-1}\right)\right]\label{errorspin}
\end{align}
We note that in our previous work~\cite{Cort} we denoted $a_{l+1}^-$ by $a_l^-$
in the asymptotic formula Eq.(\ref{eq:excitationEnergies}) which amounts to a different labeling convention
for the characteristic values. Here we stick to a more common convention. \\
For this work we need an accurate representation of $C_0 (z;q)$ for small values of $z$.
A representation that is valid for large $q$ in the interval $|z|<\pi/2$ is given by
\begin{align}
&C_0(z,q)=\frac{C_0(0,q)}{\sqrt{2}}  \nonumber \\
&\times\frac{e^{2\sqrt{q}\sin(z)}\cos\left(\frac{z}{2}+\frac{\pi}{4}\right)+e^{-2\sqrt{q}\sin(z)}\sin\left(\frac{z}{2}+\frac{\pi}{4}\right)}{\cos z}\label{MCLargeNu2}
\end{align}
To the determine this function we also need to know its prefactor $C_0 (0;q)$ which is given by~\cite{Gertrud}
\begin{equation}
C_0(0,q)=C_0\left(\frac{\pi}{2};q\right)2^{3/2}e^{-2\sqrt{q}}\left[1+\frac{1}{16 q^{1/2}}+\frac{9}{256 q}\right] 
\end{equation}
This equation involves yet another prefactor which is obtainable from Sips' expansion~\cite{Cort}
and given in leading order in $q$ to be
\begin{equation}
C_{0}(\frac{\pi}{2};q) =\left(\frac{\pi \sqrt{q}}{2 }\right)^{1/4}\left(1+\frac{1}{8 \sqrt{q}} + \frac{27}{512 q}+  .. \right)^{-1/2} .\label{norm1} 
\end{equation}
In particular we find that
\begin{equation}
C_0^2 (0,q) = 4 (2 \pi)^{1/2} q^{1/4} e^{-4 \sqrt{q} }  \quad (q \rightarrow \infty)
\label{cooo}
\end{equation}
from which we obtain the density in the bond midpoint of Eq.(\ref{eq_n0}).

\section{Analysis of the function $\mathcal{P} (x)$}

\label{AppendixB}

\begin{figure}[ht]
\centering
\includegraphics[width=0.5\textwidth]{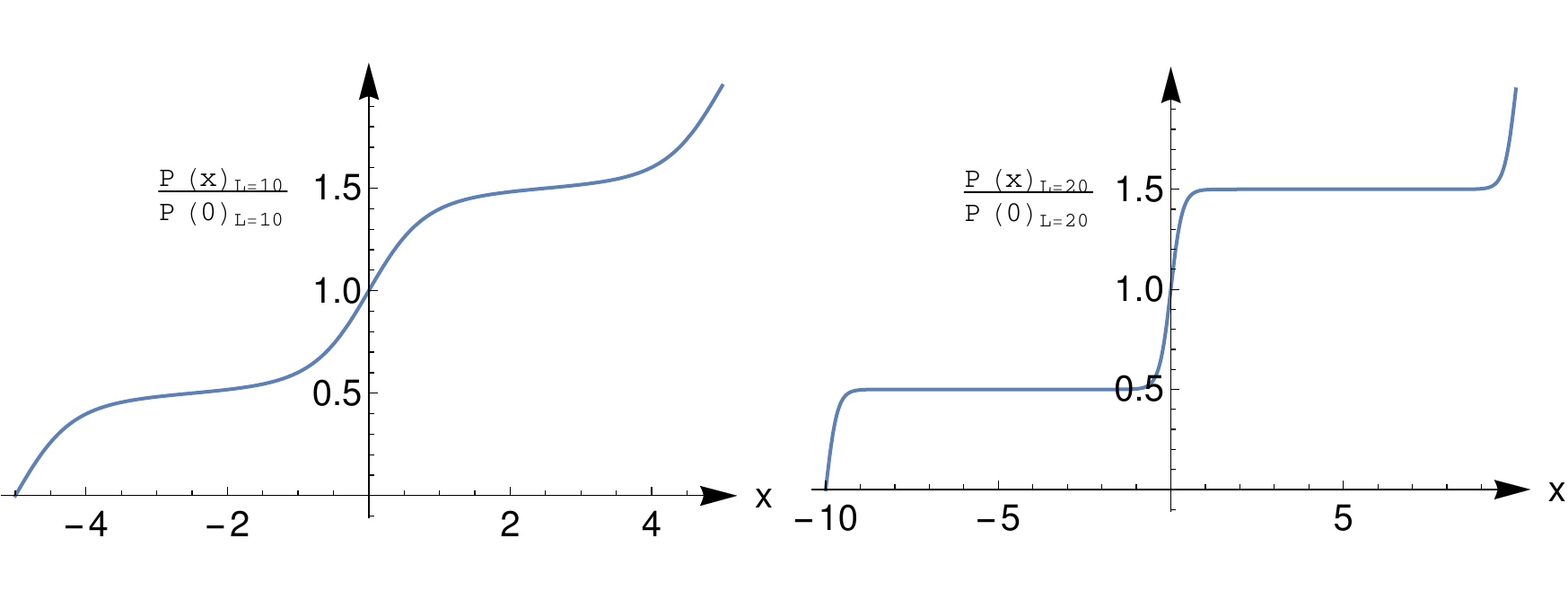}
\label{Fig10}
\caption{The function $\mP(x) /\mP(0)$ for $L=10$ (left panel) and $L=20$ (right panel) where we clearly see a step structure appearing for increasing $L$.} 
\end{figure}

We study here the properties of the function $\mathcal{P} (x)$ defined in Eq.(\ref{px_func}) we rewrite here as
\begin{align}
\mathcal{P} (x) &= w'' (\frac{L}{2}) \frac{L}{4} \int_{-L/2}^x \frac{dy}{C_0^2 (2 \pi y /L; \nu)} \nonumber \\
& = \gamma \int_{-\pi}^{2\pi x /L} f(t,\nu)  dt
\end{align}
where we used the explicit form of the density and we defined
\begin{align}
f(z,\nu) & = \frac{C_0^2 (0,\nu)}{C_0^2 (z,\nu)} \nonumber \\
\gamma &=  \frac{L^2}{8 \pi} \frac{w'' (\frac{L}{2})}{C_0^2 (0; \nu)} \label{gamma}
\end{align}
It will be convenient to further introduce the functions
\begin{align}
\mathcal{I}(z,\nu) &= \int_{0}^{z}dt\ f(t;\nu)   \label{Iz}
\end{align}
and $\mathcal{J} (\nu)  = \mathcal{I} (\pi/2,\nu) $ such that we can write
\begin{equation}
\mathcal{P} (x) =\gamma \left[  2 \mathcal{J} (\nu) + \mathcal{I} (\frac{2 \pi x}{L}, \nu)  \right]
\end{equation}
where we used the symmetry of the integrand.
Using then the asymptotic expansion of Mathieu functions functions \Eq{MCLargeNu2}, the $f(t;\nu)$ reads
\begin{align}
& f(z,\nu)= \nonumber \\ 
&\frac{2 \cos^2 z}{\left[e^{2\sqrt{\nu}\sin(z)}\cos\left(\frac{z}{2}+\frac{\pi}{4}\right)+e^{-2\sqrt{\nu}\sin(z)}\sin\left(\frac{z}{2}+\frac{\pi}{4}\right)\right]^2} 
\end{align}
For $\nu$ very large this function has its main contributions from $z=0$ and we can approximate
\begin{equation}
f(z,\nu) = \frac{\cos z}{\cosh^2 ( 2 \sqrt{\nu} \sin z)}
\end{equation}
which inserted into Eq.(\ref{Iz}) gives
\begin{equation}
I (z,\nu) = \frac{\tanh (2 \sqrt{\nu} \sin z)}{ 2 \sqrt{\nu}}
\end{equation}
and consequently
\begin{equation}
\mathcal{J} (\nu) =\frac{\tanh (2 \sqrt{\nu} )}{ 2 \sqrt{\nu}} = \frac{1}{2 \sqrt{\nu}}  \quad (\nu \rightarrow \infty) .
\end{equation}
From this we can evaluate $\mathcal{P}(0)$. Using Eq.(\ref{cooo}) and (\ref{gamma}) we find
\begin{equation}
\mathcal{P} (0) = 2\gamma \mathcal{J}(\nu) = 
w''(\frac{L}{2})  \frac{L^2}{16 (2 \pi)^{3/2} \nu^{3/4}}e^{4\sqrt{\nu}} \quad (L \rightarrow \infty)
\end{equation}
which yields Eq.(\ref{P_limit}). Finally we consider the quantity
\begin{align}
\frac{\mathcal{P} (x) - \mathcal{P} (0)}{\mathcal{P} (0)} &= \frac{\mathcal{I} (2\pi x /L,\nu)}{2 \mathcal{J} (\nu)} \nonumber \\
&= \frac{\tanh (2 \sqrt{\nu} \sin (2\pi x /L))}{2 \tanh (2 \sqrt{\nu})} \nonumber \\
&= \theta (x) - \half    \quad (L \rightarrow \infty)
\end{align}
and therefore
\begin{equation}
\mathcal{P} (x) = \mathcal{P} (0) \left[ \half + \theta (x) \right]  \quad (L \rightarrow \infty)
\end{equation}
This behavior of the function $\mathcal{P} (x)$ is illustrated in Fig.\ref{Fig10}
where we plotted $\mathcal{P}(x) /\mathcal{P} (0)$. In this figure we clearly see the step appearing with increasing $L$.

\end{document}